\documentclass[a4paper]{jpconf}
\usepackage{graphicx}
\usepackage{epsfig}
\usepackage{amssymb}
\usepackage{amsmath, amscd}
\usepackage{cite}
\newcommand{\R}{\mathbb{R}}

\newcommand{\C}{\mathbb{C}}

\usepackage{lipsum}

\newenvironment{itquote}
  {\begin{quote}\itshape}
  {\end{quote}\ignorespacesafterend}

\begin{document}
%\title{On BGS-conjecture, CLT and measurement}
\title{The measurement in classical and quantum theory}

\author{Alexey A. Kryukov}

\address{Department of Mathematics \& Natural Sciences, University of Wisconsin-Milwaukee, USA}

\ead{kryukov@uwm.edu}

\begin{abstract}
The Bohigas-Giannoni-Schmit (BGS) conjecture states that the Hamiltonian of a microscopic analogue of a classical chaotic system can be modeled by a random matrix from a Gaussian ensemble. Here, this conjecture is considered in the context of a recently discovered geometric relationship between classical and quantum mechanics. 
%Based on
Motivated by BGS, we conjecture that the Hamiltonian of a system whose classical counterpart performs a random walk can be modeled by a family of independent random matrices from the Gaussian unitary ensemble. By accepting this conjecture, we find a relationship between the process of observation in classical and quantum physics, derive irreversibility of observation 
and describe the boundary between the micro and macro worlds.
% and provide a rigorous and non-contradictory transition between classical and quantum domains. 
\end{abstract}

% Edits: 1. A new intro section is written, where the questions asked by the referees are answered in the form of an introduction.
%  2. To make all statements in the paper be logically connected, the proofs of all statements are provided and the logic connecting the theorems is highlighted. This resulted in a more direct theorem-proof format with additional clarifications offered in between. 
% Eliminated several particle discussion to avoid the issue of applicability of BGS to many particle systems

\section{Introduction}

Given the fundamental character of the topics discussed in the paper, it is necessary to place the paper in the context of existing research and to explain prerequisites for the obtained results. First and foremost, standard quantum mechanics is assumed throughout the paper. The equation of motion is the usual Schr{\"o}dinger equation, with no extra terms. In proving the initial results related to the connection of Schr{\"o}dinger and Newtonian dynamics, the state of the system that represents a macroscopic particle is assumed to be constrained to a certain submanifold in the space of states. This may seem to contradict the Schr{\"o}dinger dynamics of the system. However, this step is only used to understand the geometry of the constraint and is later replaced by a condition on the Hamiltonian of the system. Second, the presented results are mathematical theorems proved in the paper and therefore, given the assumptions in the theorems, the results follow rigorously and need not be questioned, unless the validity of the proof itself is in question. What is open to discussion is the nature of the assumptions and the physical interpretation of the derived results.

Now, the only assumption of essence used in proving the results is related to a specific form of the Hamiltonian of a macroscopic particle in natural surroundings and a microscopic particle under a measurement. This assumption is stated in the conjecture ${\bf (RM)}$. It claims that the Hamiltonian of a microscopic particle whose position is measured, as well as a macroscopic particle in the natural surroundings can be represented by a random matrix from a Gaussian unitary ensemble and that random matrices considered at different moments of time are independent. 
Of course, ${\bf (RM)}$ is not BGS. In particular, BGS is about stationary states and the distribution of eigenvalues of the Hamiltonian. On the contrary, ${\bf (RM)}$ is about a non-stationary stochastic process on the space of states. 
At the same time, both conjectures claim that under certain physical conditions on the system the Hamiltonian of the system can be represented by a random matrix.
 This is the point of connection of the paper with the BGS conjecture. It is important to point out that there is no attempt to prove or disprove the BGS conjecture, or to justify it in any way in the paper. Rather, the validity of the BGS conjecture confirmed in numerous observations is used to argue in favor of ${\bf (RM)}$. One may disagree with the argument, but given that ${\bf (RM)}$ is the only assumption of essence used in proving the theorems in the paper, and the statements of theorems have far reaching consequences to physics, the approach deserves our full attention.

Finally, what are the results obtained in the paper?  How are they related to previous research?  What is the main merit of the paper? 
%How potentially significant are they?
%what are the results, how are they related to previous research, and how potentially significant are they? 
The coherent states used in identifying the classical phase space manifold in the paper are well known and widely used in quantum optics and quantum physics in general. The fact that when the state is coherent, the motion of the system is closest to the classical motion and, in the limit, may become identical to it is not surprising. The literature related to the subject ranges from textbook results in quantum optics, e.g., \cite{Klauder} to research in quantum gravity (see \cite{Martin} for a historic context, main ideas and some references). A semiclassical approximation obtained by constraining the state to remain Gaussian during the evolution \cite{Heller} is known to be useful in modeling a range of quantum systems over a time interval. 

However, the current paper is not concerned with a semiclassical approximation. It deals with the full-fledged Schr{\"o}dinger evolution with the Hamiltonian that satisfies ${\bf (RM)}$. The state of the system under such evolution does not stay coherent, but performs a random walk in the space of states. This walk is used in the paper to find a
%
%The main results in the paper concern the 
new link between measurement in classical and quantum mechanics. 
%Assuming ${\bf (RM)}$, it becomes possible to model the process of observation on macroscopic and microscopic particles alike. 
%
Namely, the normal distribution typical for classical observations and the Born rule valid in the micro-world turn out to follow from the Schr{\"o}dinger equation with the Hamiltonian satisfying ${\bf (RM)}$. It becomes possible to specify on this basis the boundary between the micro and macro worlds. It becomes also possible to explain irreversibility of measurement. Other essential results are the derivation of the classical behavior of macroscopic systems and an explanation of why the cat states of macroscopic systems are not observed in nature. This in turn provides an explanation of the role played by measuring devices in quantum theory and how the quantum theory, while remaining linear, may avoid using superpositions of states of macroscopic bodies.

As stated before, the derived results are mathematical consequences of the assumption ${\bf (RM)}$. They are new and are not directly based on previous research. 
%The literature referenced in the paper serves only to provide a background and historic information. 
%
The measurement problem and physics of measuring devices is of course a topic of significant research of its own. However, as far as the author is aware, the existing body of literature related to quantum measurement was never concerned with ${\bf (RM)}$. 
The theory of decoherence \cite{Decoh} probably came closest  and is possibly  most relevant to the topics considered in the paper. However, in classification of the decoherence theory, we are dealing here with a ``fake" decoherence, as the evolution remains unitary. Furthermore, the paper deals with evolution of the state and not the density matrix of a system. The irreversibility of a measurement is the result of the Hamiltonian lacking time-reversal symmetry rather than a ``leakage" of information into the environment.
%Therefore, it does not seem to be relevant to refer to research in decoherence theory in the paper. 
Similarly, the dynamical collapse theories surveyed in \cite{Bassi1, Bassi2} use non-linear modifications of the Schr{\"o}dinger equation, while only linear evolution is considered in the paper. Accordingly, we only refer to general surveys in both of these fields of research with the goal of providing a context for the derived results.
Let us mention the Ornstein-Uhlenbeck process of stochastic relaxation of Hermitian matrices first considered by Dyson \cite{Dyson, Fyodorov}. Although not directly related to the stochastic process considered here, it provides an interesting example of a process that leads to the Gaussian ensemble used in the paper.

To make the presentation self-contained, simple proofs of the previously published theorems {\bf (A)} and {\bf (B)}, and propositions {\bf (P2)} and {\bf (P3)}  are provided. The reader is referred to the cited publications for details.

\section{The measurement in classical and quantum theory}

The linear nature of quantum mechanics poses a persistent problem for reconciliation of the classical and quantum mechanics. The superposition principle is foreign to classical physics. We don't see macroscopic objects in two different places or the cat being alive and dead at the same time. However, such states are commonplace in the microworld. The notable attempts to resolve the situation include non-linear modifications of the Schr{\"o}dinger equation to account for the transition to states observed in a measurement, the De Broglie-Bohm theory of classical particles lead by a pilot-wave, an appeal to many coexisting worlds representing the components of a superposition and the decoherence program aiming to explain how superpositions decohere to probabilistic mixtures. None of these attempts is generally accepted as successful in resolving the problem. There is no experimental evidence for needing to modify the Schr{\"o}dinger equation. The simultaneous presence of quantum and classical trajectories in the pilot-wave theory seems redundant. The many-world approach fails to explain the world that is real and unique to us. The decoherence program derives the laws of probability valid for macroscopic bodies but fails to account for a specific outcome of a measurement. 

On the other hand, the Newtonian and Schr{\"o}dinger dynamics have a simple relationship that does not seem to be known nor used in the above-mentioned attempts. Namely, the Newtonian dynamics can be identified with a constrained Schr{\"o}dinger dynamics. The latter is similar to the Newtonian dynamics of a constrained system, e.g., a bead on a wire. However, since the Schr{\"o}dinger equation describes the dynamics in a Hilbert space of states, the constraint is imposed on the state of the system. 
Consider the subset $M^{\sigma}_{3,3}$ of the Hilbert space $L_{2}(\R^3)$, formed by the states 
\begin{equation}
\varphi({\bf x})=g_{{\bf a}, \sigma}({\bf x})e^{i{\bf p}{\bf x}/\hbar},
\end{equation} 
where 
\begin{equation}
g_{{\bf a}, \sigma}=\left(\frac{1}{2\pi\sigma^{2}}\right)^{3/4}e^{-\tfrac{({\bf x}-{\bf a})^{2}}{4\sigma^{2}}}
\end{equation}
is the Gaussian function of a sufficiently small variance $2\sigma^2$ centered at a point ${\bf a}$ in the Euclidean space $\R^3$ and ${\bf p}$ is a fixed vector in $\R^3$. The state $\varphi({\bf x})$ represents a narrow wave packet with group velocity ${\bf p}/m$, where $m$ is the mass of the particle.
Let's identify the set of all pairs $({\bf a}, {\bf p})$ with the classical phase space $\R^3 \times \R^3$ of possible positions ${\bf a}$ and momenta ${\bf p}$ of a particle. The map $\Omega: ({\bf a}, {\bf p}) \longrightarrow g_{{\bf a}, \sigma}e^{i{\bf p}{\bf x}/\hbar}$ identifies then the classical phase space with the submanifold  $M^{\sigma}_{3,3}$ of $L_{2}(\R^3)$. The equivalence classes of states in $L_{2}(\R^3)$ differing only by a constant phase factor $e^{i\alpha}$ form the projective space $CP^{L_{2}}$. 
Under the equivalence relation, the embedded manifold $M^{\sigma}_{3,3}$ becomes a submanifold of $CP^{L_{2}}$, denoted here by the same symbol. 

The six-dimensional manifold $M^{\sigma}_{3,3}$ is embedded into the space of states in a very special way. There are no vectors in the Hilbert space $L_{2}(\R^3)$ orthogonal to all of $M^{\sigma}_{3,3}$. Instead, the points of $M^{\sigma}_{3,3}$ represent an overcomplete basis in the Hilbert space \cite{KryukovMath}. Furthermore, the projective space $CP^{L_{2}}$ possesses the Fubini-Study metric, induced by the embedding of $CP^{L_{2}}$ into the sphere $S^{L_{2}}$ of unit-normalized states in $L_{2}(\R^3)$ furnished itself with the induced round Riemannian metric. For any two vectors $\xi, \eta$ in $L_{2}(\R^3)$ tangent to the sphere $S^{L_{2}}$ and the corresponding vectors $X=(\mathrm{Re} \xi, \mathrm{Im} \xi)$ and  $Y=(\mathrm{Re} \eta, \mathrm{Im} \eta)$ in the realization $L_{2R}(\R^3)$ of $L_{2}(\R^3)$,   the Riemannian metric $G_{\varphi}$ on the sphere is defined by
\begin{equation}
G_{\varphi}(X,Y)=\mathrm{Re}(\xi,\eta).
\end{equation}
 Perhaps surprisingly, the metric induced on the submanifold  $M^{\sigma}_{3,3}$ of $CP^{L_{2}}$ with $2\sigma$ as a unit of length turns out to be the ordinary Euclidean metric. In other words, the map $\Omega: \R^3 \times \R^3 \longrightarrow CP^{L_{2}}$ is an isometric embedding of the Euclidean space into the space of states. The manifold $M^{\sigma}_{3,3}$ can be also furnished with a compatible linear structure, making it isomorphic to  the Euclidean space $\R^3 \times \R^3$ \cite{KryukovMacro, KryukovPosition, Kryukov2020, KryukovMath, KryukovUncert}.
 
 %%%%  Insert 2
 
 We can now state the first theorem. Here and later, it is assumed that the value of the parameter $\sigma$ is sufficiently small for the linear approximation of the potential $V({\bf x},t)$ to be valid within intervals of length $\sigma$.
\begin{itquote}{\bf{(A)}}
The Newtonian dynamics of an arbitrary mechanical system is the Schr{\"o}dinger dynamics of that system with the state constrained to the classical phase space submanifold of the space of states of the system. Furthermore, the Schr{\"o}dinger dynamics is the only unitary evolution that reduces under the constraint to the Newtonian one.
\end{itquote}

Note that for the purpose of this paper we only need the proof of the first part of the theorem and only in the case of a single particle. A complete prove can be found in \cite{Kryukov2020}.

{\it Proof.} A simple way to prove 
%that Newtonian dynamics of a particle is a constrained Schr{\"o}dinger dynamics 
the theorem is by using the variational principle.
 %%%%%  End Insert 2 
%
The variation of the functional 
\begin{equation}
\label{SS}
S[\varphi]=\int \overline{\varphi}({\bf x},t) \left[i\hbar \frac{\partial}{\partial t}-{\widehat h}\right] \varphi({\bf x},t) d^3 {\bf x} dt
\end{equation}
with the Hamiltonian ${\widehat h}=-\frac{\hbar^{2}}{2m}\Delta+V({\bf x},t)$ yields the Schr{\"o}dinger equation for $\varphi$. For the states $\varphi$ constrained to the manifold $M^{\sigma}_{3,3}$, this functional reduces to the classical action
\begin{equation}
S=\int \left[{\bf p}\frac{d {\bf a}}{dt}-h({\bf p},{\bf a},t)\right]dt,
\end{equation}
where $h({\bf p},{\bf a},t)=\frac{{\bf p}^2}{2m}+V({\bf a},t)+const$ is the Hamiltonian function for the particle. We used here the fact that for a sufficiently small $\sigma$, the terms $V({\bf a},t)$ and $\int g^2_{{\bf a}, \sigma}({\bf x})V({\bf x},t)d^3 {\bf x}$ are arbitrarily close to each other. In fact, as $\sigma$ approaches $0$, the terms $g^2_{{\bf a}, \sigma}$ form a delta sequence.
It follows that the variation of the functional (\ref{SS}) subject to the constraint that $\varphi$ belongs to $M^{\sigma}_{3,3}$ yields Newtonian equations of motion.
\hfill$\square$

%Insert 3

The theorem provides us with a geometric condition for the transition from Schr{\"o}dinger to Newtonian dynamics. Note that we are not trying to change the Schr{\"o}dinger equation by imposing the constraint. Rather, the upcoming theorems will demonstrate that the constraint itself can be traced back to the Schr{\"o}dinger dynamics with a proper Hamiltonian. Theorem $\bf{(A)}$ will play an important role in this derivation. To proceed, let us first work on a deeper understanding of geometry and physics behind the theorem. 
%Insert 3a
The reader who is more interested in the applications of the theorem, may skip propositions {\bf (1)}  and {\bf (2)} below.
%End of Insert 3a

{\bf (1)} The transition from Schr{\"o}dinger to Newtonian dynamics can be expressed in terms of transition from quantum commutators to Poisson brackets. 
%This complements the results announced in \cite{Kryukov2020}. 
A simple calculation with the above $h({\bf p},{\bf a},t)$ and ${\widehat h}$, and $\varphi$ in $M^{\sigma}_{3,3}$ yields the following:
%
%End Insert 3
%
\begin{equation}
\label{aa}
(\varphi, \tfrac{1}{i\hbar}[{\widehat {\bf x}}, {\widehat h}]\varphi)= \{{\bf a}, h \},
\end{equation}
and
\begin{equation}
\label{bb}
(\varphi, \tfrac{1}{i\hbar}[{\widehat {\bf p}}, {\widehat h}]\varphi)=\{{\bf p}, h\}.
\end{equation}
The brackets on the right side of (\ref{aa}) and (\ref{bb}) are the usual Poisson brackets. Consider the linear vector fields $x_{\varphi}={\widehat {\bf x}}\varphi$ and $p_{\varphi}={\widehat {\bf p}}\varphi$ on $L_{2}(\R^3)$ associated with the operators of position and momentum. For a given state $\varphi$, let $\overline{\bf x}$ and $\overline{\bf p}$ be the expected values of the operators on this state. The components $x_{\varphi \perp}=({\widehat {\bf x}}-\overline{\bf x} I)\varphi$ and  $p_{\varphi \perp}=({\widehat {\bf p}}-\overline{\bf p} I)\varphi$ are tangent to the sphere $S^{L_{2}}$ at $\varphi$ and orthogonal to the fibres $\{\varphi \}$ of the fibre bundle  $CP^{L_{2}}=S^{L_{2}}/S^1$. Therefore, they are tangent to the projective space $CP^{L_{2}}$ itself. Moreover, the vector fields  $x_{\varphi \perp}$ and  $p_{\varphi \perp}$ constrained to the submanifold $M^{\sigma}_{3,3}$ of $CP^{L_{2}}$ are tangent to it. The integral curves of these fields are lines of constant position and momentum that provide the coordinate grid in the classical phase space $\R^3 \times \R^3$. This can be seen directly by comparing the right hand sides of the equations
\begin{equation}
({\widehat {\bf x}}-\overline{\bf x}I)\varphi=({\bf x}-{\bf a})\varphi
\end{equation}
and
\begin{equation}
({\widehat {\bf p}}-\overline{\bf p}I)\varphi=-\frac{i\hbar}{2\sigma^2}({\bf x}-{\bf a})\varphi,
\end{equation}
valid for $\varphi$ in $M^{\sigma}_{3,3}$, with the gradients $\nabla_{\bf a} \varphi$ and $\nabla_{\bf p}\varphi$. The latter gradients represent vectors tangent to the lines of constant values of ${\bf p}$ and ${\bf a}$  through a point of the projective manifold $M^{\sigma}_{3,3}$.

%Insert 4
%Therefore, we obtain the following restatement of the variational principle for the functional (\ref{SS}) with the state $\varphi$ constrained to manifold $M^{\sigma}_{3,3}$:
We have the following proposition.
\begin{itquote}{{\bf (P1)}}
At any point $({\bf a}, {\bf p})$ of the phase space $\R^3\times \R^3$, the derivatives $\frac{d{\bf a}}{dt}$ and $\frac{d{\bf p}}{dt}$ of a Newtonian motion are projections at the point $\Omega({\bf a}, {\bf p})$  of the velocity $\frac{d\varphi}{dt}$ of the Schr{\"o}dinger evolution onto the coordinate lines of the phase space submanifold $M^{\sigma}_{3,3}$. 
\end{itquote}
%
%End Insert 4

{\it Proof.} Using the Ehrenfest theorem written in terms of the vector fields $x_{\varphi \perp}$ and $p_{\varphi \perp}$ and (\ref{aa}) and (\ref{bb}), we have for the initial state $\varphi$ in $M^{\sigma}_{3,3}$ at $t=0$:
\begin{equation}
\label{EE1}
2\mathrm{Re} \left(\frac{d\varphi}{dt}, x_{\varphi \perp}\right)
%=(\varphi, \tfrac{1}{i\hbar}[{\widehat {\bf x}}, {\widehat h}]\varphi)
=\{{\bf a}, h\}=\frac{d{\bf a}}{dt}
\end{equation}
and
\begin{equation}
\label{EE2}
2\mathrm{Re} \left(\frac{d\varphi}{dt}, p_{\varphi \perp} \right)
%=(\varphi, \tfrac{1}{i\hbar}[{\widehat {\bf p}}, {\widehat h}]\varphi)
=\{{\bf p}, h\}=\frac{d{\bf p}}{dt}.
%=- \nabla V({\bf a}).
\end{equation}
\hfill$\square$

{\bf (2)} Furthermore, it is possible to decompose the initial velocity of state of a particle in a potential $V$ with the initial state in $M^{\sigma}_{3,3}$ into four physically meaningful orthogonal components. The two components tangent to the phase space submanifold $M^{\sigma}_{3,3}$ were already identified with the usual classical velocity ${\bf v}=d{\bf a}/dt$ and acceleration ${\bf w}=d{\bf v}/dt=-\nabla V/m$ of the particle. The other two components orthogonal to $M^{\sigma}_{3,3}$ are the {\em phase velocity} (projection of $d\varphi/dt$ onto the unit vector $-i\varphi$) and the {\em velocity of spreading} of the wave-packet  (projection of $d\varphi/dt$ onto the unit vector in the direction $id\varphi/d\sigma$). 
\begin{itquote}{{\bf (P2)}}
The norm of the total velocity $d \varphi/dt$ at $t=0$, i.e., the speed of motion of the initial state $\varphi$, is given by the formula
\begin{equation}
\label{decomposition}
\left\|\frac{d\varphi}{dt}\right\|^{2}=\frac{{\overline E}^{2}}{\hbar^{2}}+\frac{{\bf v}^{2}}{4\sigma^{2}}+\frac{m^{2}{\bf w}^{2}{\sigma}^{2}}{\hbar^{2}}+\frac{\hbar^{2}}{32\sigma^{4}m^{2}}.
\end{equation}
When the state is constrained to the manifold $M^{\sigma}_{3,3}$, the first term (the phase velocity squared) and the last term (the velocity of spreading squared) disappear. The only surviving terms are associated with the classical velocity and acceleration and the motion follows the classical Newtonian dynamics in the phase space.
\end{itquote}
%Insert 5
{\it Proof.} The decomposition (\ref{decomposition}) is obtained by projecting the velocity $\frac{d \varphi}{dt}$ of state under the Schr{\"o}dinger evolution with an arbitrary Hamiltonian ${\widehat h}=-\frac{\hbar^{2}}{2m}\Delta+V({\bf x},t)$ at a point $\varphi=g_{{\bf a}, \sigma}e^{i{\bf p}{\bf x}/\hbar}$ in the classical phase space $M^{\sigma}_{3,3}$ onto an orthonormal set of vectors specified by changing the parameters, ${\bf a}, {\bf p}, \sigma$ that define $\{\varphi\}$ and the phase parameter $\theta$ of a possible constant phase factor $e^{-i\theta}$ of $\varphi$. Calculation of the classical space components of $\frac{d \varphi}{dt}$  at an arbitrary point $\varphi$ in $M^{\sigma}_{3,3}$ yields
\begin{equation}
\label{pproj}
\left.\mathrm{Re}\left(\frac{d \varphi}{dt}, -\widehat{\frac{\partial \varphi}{\partial a^{\alpha}}}\right)\right|_{t=0}=\frac{p^{\alpha}}{2m\sigma},
\end{equation}
where the hat here and in other calculations denotes normalization.
For the momentum space components of $\frac{d \varphi}{dt}$  at $\varphi$  we similarly obtain, assuming that $\sigma$ is small enough to make the linear approximation of $V({\bf x})$ valid:
\begin{equation}
\label{pproj1}
\left.\mathrm{Re}\left(\frac{d \varphi}{dt}, \widehat{\frac{\partial \varphi}{\partial p^{\alpha}}}\right)\right|_{t=0}=\frac{mw^{\alpha} \sigma}{\hbar}, \quad {\text where} \quad
mw^{\alpha}=-\left.\frac{\partial V({\bf x})}{\partial x^{\alpha}}\right|_{{\bf x}={\bf a}}.
\end{equation}
The components (\ref{pproj}) and (\ref{pproj1}) are tangent to $M^{\sigma}_{3,3}$ and orthogonal to the fibre $\{\varphi\}$.
The component of the velocity $\frac{d\varphi}{dt}$ due to change in $\sigma$ (spreading)  is orthogonal to the phase space $M^{\sigma}_{3,3}$ and the fibre $\{\varphi\}$, and is equal to
\begin{equation}
\label{spreadcomp}
\mathrm{Re} \left (\frac{d\varphi}{dt}, i\widehat{\frac{d\varphi}{d\sigma}}\right)=\frac{\sqrt{2}\hbar}{8\sigma^{2}m}.
\end{equation}
The component of the velocity parallel to the fibre $\{\varphi\}$ is the expected value of energy divided by $\hbar$:
\begin{equation}
\label{phase}
\mathrm{Re} \left (\frac{d\varphi}{dt}, -\widehat{\frac{d\varphi}{d\theta}}\right)=\frac{1}{\hbar}(i{\widehat h}\varphi, i\varphi)=\frac{{\overline E}}{\hbar}.
\end{equation}
Calculation of the norm of $\frac{d\varphi}{dt}=\frac{i}{\hbar}{\widehat h}\varphi$ at $t=0$ gives
\begin{equation}
\label{decompositionP}
\left\|\frac{d\varphi}{dt}\right\|^{2}=\frac{{\overline E}^{2}}{\hbar^{2}}+\frac{{\bf p}^{2}}{4m^2\sigma^{2}}+\frac{m^{2}{\bf w}^{2}{\sigma}^{2}}{\hbar^{2}}+\frac{\hbar^{2}}{32\sigma^{4}m^{2}},
\end{equation}
which is the sum of squares of the found components. This completes a decomposition of the velocity of state at any point $\varphi$ in  $M^{\sigma}_{3,3}$. 
%EndInsert5
\hfill$\square$

 Similar results are true for systems of $n$ particles \cite{Kryukov2020}. In this case, the transition to classical Newtonian dynamics happens when the state  of the system is constrained to the phase space submanifold $M^{\sigma}_{3n,3n}= M^{\sigma}_{3,3} \otimes ... \otimes M^{\sigma}_{3,3}$ in the space of states of the system. That is, the constrained state is the product of states of the $n$ particles in the system with the state of each particle constrained to a copy of $M^{\sigma}_{3,3} $.  
 %Insert 6
 Although not directly used in the paper, the results {\bf (1)} and {\bf (2)} provide an insight into the theorem ${\bf (A)}$ and help a deeper understanding of the upcoming theorems.
 %end Insert 6
 
 %%% Cut
 
 %A more detailed analysis leads to the following theorem \cite{KryukovPosition}:
%\begin{itquote}{\bf{(A)}}
%The Newtonian dynamics of an arbitrary mechanical system is the Schr{\"o}dinger dynamics of that system with the state constrained to the classical phase space submanifold of the space of states of the system. Furthermore, the Schr{\"o}dinger dynamics is the only unitary evolution that reduces under the constraint to the Newtonian one.
%\end{itquote}
%We note that the relationship between commutators and Poisson brackets established in (\ref{aa}) and (\ref{bb}) generalizes directly to systems of $n$ particles. 

%%% End cut

{\bf (3)} The identification of the classical phase space of a particle with the manifold $M^{\sigma}_{3,3}$ yields a useful metric relationship. Let $M^{\sigma}_{3}$ denote the submanifold of $M^{\sigma}_{3,3}$ of the Gaussian states $g_{{\bf a}, \sigma}=\left(\frac{1}{2\pi\sigma^{2}}\right)^{3/4}e^{-\tfrac{({\bf x}-{\bf a})^{2}}{4\sigma^{2}}}$. The map $\omega: {\bf a} \longrightarrow g_{{\bf a}, \sigma}$, $\omega(\R^3)=M^{\sigma}_{3}$ identifies the submanifold $M^{\sigma}_{3}$ of $CP^{L_{2}}$ with the classical space $\R^3$. We then have the following proposition.

\begin{itquote}{{\bf (P3)}}
Let $\theta(g_{{\bf a}, \sigma}, g_{{\bf b}, \sigma})$ be the distance from $g_{{\bf a}, \sigma}$ to  $g_{{\bf b}, \sigma}$ in $M^{\sigma}_{3}$ in the Fubini-Study metric on the space of states $CP^{L_{2}}$ and let $({\bf a}-{\bf b})^{2}$ be the square of the Euclidean distance between the corresponding points ${\bf a}$ and ${\bf b}$ in $\R^3$. Then
\begin{equation}
\label{mainO}
e^{-\frac{({\bf a}-{\bf b})^{2}}{4\sigma^{2}}}=\cos^{2}\theta(g_{{\bf a}, \sigma}, g_{{\bf b}, \sigma}).
\end{equation}
Likewise, for arbitrary states $\varphi=g_{{\bf a}, \sigma}e^{i{\bf p}{\bf x}/\hbar}$ and $\psi=g_{{\bf b}, \sigma}e^{i{\bf q}{\bf x}/\hbar}$ in the classical phase space $M^{\sigma}_{3,3}$, the distances in the Euclidean phase space and the space of states are related by 
\begin{equation}
\label{mainOG}
e^{-\frac{({\bf a}-{\bf b})^{2}}{4\sigma^{2}}-\frac{({\bf p}-{\bf q})^{2}}{\hbar^2/\sigma^{2}}}=\cos^{2}\theta(\varphi, \psi).
\end{equation}
\end{itquote}
%The derivation of (\ref{mainO}) and (\ref{mainOG}) is explained in the Appendix.
%Insert 7
 {\it Proof.} The metric relationship (\ref{mainO}) follows from the inner product of two states in $M^{\sigma}_{3}$:
 \begin{equation}
 \label{mainOP}
 (g_{{\bf a}, \sigma}, g_{{\bf b}, \sigma})=e^{-\frac{({\bf a}-{\bf b})^{2}}{8\sigma^{2}}}.
 \end{equation}
 This expression squared is equal to the right hand side of (\ref{mainO}) by the definition of the Fubini-Study distance between states in $CP^{L_{2}}$. The result (\ref{mainOG}) is obtained in a similar way by evaluating the Fourier transform of  a Gaussian function along the way.
%End Insert 7
\hfill$\square$

The relation (\ref{mainO}) implies an important connection between probability distributions of multivariate random variables valued in the classical space and the space of states. 
Namely, consider a random variable $\varphi$ with values in the space of states $CP^{L_{2}}$.  
Because the classical space $M^{\sigma}_{3}$ is a submanifold of $CP^{L_{2}}$, we can restrict $\varphi$ to take values in $M^{\sigma}_{3}$. We then have the following theorem.
\begin{itquote}{\bf{(B)}}
Suppose the conditional probability of $\varphi$ given that $\varphi$ is in the classical space is described by the normal distribution. Suppose the probability $P$ of transition between any two states depends only on the distance between the states. Then $P$ is given by the Born rule, i.e., $P= |(\varphi, \varphi_0)|^2$, where $\varphi_0$ is the initial and $\varphi$ is the observed state. The opposite is also true: For states in the classical space the Born rule yields the normal distribution of the observed position.
\end{itquote}
So, the normal probability distribution is the Born rule in disguise! 
%Insert 8

{\it Proof.} In fact, (\ref{mainO}) establishes the needed connection between the normal distribution and the Born rule for the states in $M^{\sigma}_{3}$. By assumption, the probability of transition between states depends only on the Fubini-Study distance between them. Since the Fubini-Study distance between states in $M^{\sigma}_{3}$ takes on all possible values, the validity of the Born rule for the states in $M^{\sigma}_{3}$ signifies its validity for arbitrary states. See \cite{KryukovPosition} for details.
%End Insert 8
\hfill$\square$

These results establish a deep connection between Newtonian and Schr{\"o}dinger dynamics and between the classical space and classical phase space, and the submanifolds $M^{\sigma}_{3}$ and $M^{\sigma}_{3,3}$ of the space of states. Based on these results we put forward the following embedding hypothesis:
\begin{itquote}{\bf{(EH)}}
The constructed mathematical embedding of the classical space and classical phase space into the space of states and the resulting identification of Newtonian dynamics with the constrained Schr{\"o}dinger dynamics are physical. That is, the classical space, phase space and the Newtonian dynamics of a system are not only fully derivable from, but also physically originate in the Schr{\"o}dinger dynamics in the Hilbert space of states of the system.
\end{itquote}
The hypothesis asserts that the provided mathematical derivation of the classical Newtonian dynamics from the Schr{\"o}dinger dynamics of the system is the true physical correspondence between classical and quantum systems. 

%Insert 9
Let us point out that this embedding hypothesis is not required for validity of the theorems in the paper. For this, the mathematical isometric embedding of $M^{\sigma}_{3,3}$ into $CP^{L_{2}}$ is sufficient. Viewed this way, the results that will be derived can be considered curious mathematical consequences of the already verified isometry of the embedding and a single forthcoming conjecture ${\bf (RM)}$. The next few paragraphs preceding ${\bf (RM)}$ can be then looked at as a physical motivation of the conjecture. Whatever point of view is preferred by the reader, the results will provide an advance in our understanding of the transition from quantum to classical theory and of the process of measurement.

To validate the hypothesis, or to motivate the forthcoming conjecture, let us show that the hypothesis 
%End Insert 9
is consistent with all observed classical and quantum phenomena. This involves showing that all observable classical phenomena can be derived from the corresponding quantum phenomena by constraining the state to the classical phase space submanifold in the space of states. This also involves physically explaining the constraint itself. Now, an arbitrary deterministic classical motion is described by the Newtonian equations of motion. The theorem ${\bf (A)}$ demonstrates that these equations follow from the Schr{\"o}dinger equation with the constraint. Therefore, given the constraint, the hypothesis is consistent with an arbitrary deterministic classical motion. To validate the hypothesis, it remains then to: (1) verify its consistency for the motion of particles described in statistical mechanics, specifically, for the Brownian motion, and (2) explain the origin of the constraint 
%Insert 10
from the Schr{\"o}dinger dynamics itself. 
%End Insert 10
The rest of the paper will be dedicated to these two tasks.

Instead of diving into the subject of a quantum version of the Brownian motion, note that a macroscopic Brownian particle in a medium can be considered a classical chaotic system. For instance, the motion of the particle through an appropriate lattice of round obstacles provides a classical chaotic realization of the Brownian motion \cite{Szasz}. A lively discussion of the stochastic, deterministic chaotic, and regular characterizations of the Brownian motion can be found in \cite{Nature1, Nature2, Cecconi}. With the view that the Brownian particle in a medium is a chaotic system comes the applicability of the BGS-conjecture to the system \cite{BGS}. It asserts that the Hamiltonian of the corresponding quantum system can be represented by a random matrix. 

Random matrices were originally introduced into quantum mechanics by Wigner \cite{Wigner} in a study of excitation spectra of heavy nuclei. Wigner reasoned that the motion of nucleons within the nucleus is so complex that the Hamiltonian of the system can be modeled by a random matrix from an ensemble that respects symmetries of the system but otherwise contains no additional information. 
It was later discovered that correlations in the spectrum of random matrices possess a remarkable universality in being applicable to a large number of quantum systems. That includes nuclear, atomic, molecular and scattering systems, chaotic systems with few degrees of freedom as well as complex systems, such as solid state systems and lattice systems in the field theory.
This wealth of experimental evidence suggests that all quantum systems whose classical counterpart is chaotic exhibit random matrix statistics. This is the essence of the BGS-conjecture.

As mentioned, from the possibility to attribute chaotic character to Brownian motion and the BGS conjecture it follows that the Hamiltonian of the quantum analogue of the Brownian motion at any time is given by a random matrix. On the physical grounds, we can also claim that the random matrices that represent the Hamiltonian at two different moments of time must be independent. This leads us to the following 
%version of the BGS-
conjecture:
\begin{itquote}{\bf{(RM)}}
The quantum-mechanical analogue of the Brownian motion can be modeled by a random walk of state in the space of states of the system. The steps of the random walk without drift satisfy the Schr{\"o}dinger equation with the Hamiltonian represented by a random matrix from the Gaussian unitary ensemble (GUE). The matrices representing the Hamiltonian at two different times are independent and belong to the same ensemble.
\end{itquote}

Note that the Schr{\"o}dinger equation with the Hamiltonian represented by a random matrix describes evolution of the state and not of the density matrix for the system. In that sense, it is analogous to the Langevin equation for the position of a Brownian particle rather than the diffusion equation for the probability density function. 
%Insert 11
Note also, that even though both, $\bf{(RM)}$ and BGS conjectures deal with random matrices, they should not be confused with each other. In particular, while BGS deals with spectra of Hamiltonians, $\bf{(RM)}$ defines a non-stationary stochastic process. A mathematically inclined reader may skip the motivation for $\bf{(RM)}$ and consider it an assumption used in the forthcoming theorems. In fact, this is the only important assumption needed for validity of the theorems in the paper.

%End Insert 11

First of all, let us prove that the random walk in  ${\bf (RM)}$ yields the Brownian motion in $\R^3$. More precisely, we have the following theorem.
\begin{itquote}{\bf{(C)}}
The random walk described in ${\bf (RM)}$ but conditioned to stay on the submanifold $M^{\sigma}_{3}$ in the space of states yields a random walk on $\R^3$ that approximates the Brownian motion of a particle in a medium. 
%Under the constraint that the state belongs to the manifold $M^{\sigma}_{3}$ and with the use of conditional probabilities, the Schr{\"o}dinger equation with the Hamiltonian described in ${\bf (RM)}$ yields a random walk on $\R^3$ that approximates the Brownian motion of the particle. 
\end{itquote}
{\it Proof.} In fact, a general Schr{\"o}dinger evolution with Hamiltonian ${\widehat h}$ can be thought of as a sequence of steps connecting the points $\varphi_{t_{0}}, \varphi_{t_{1}}, . . . $ in the space of states. For small time intervals $\Delta t=t_{k}-t_{k-1}$, the state $\varphi_{t_{N}}$ at time $t_N$ is given by the time ordered product
\begin{equation}
\label{tN}
\varphi_{t_{N}}=e^{-\frac{i}{\hbar}{\widehat h}(t_N)\Delta t}e^{-\frac{i}{\hbar}{\widehat h}(t_{N-1})\Delta t}... e^{-\frac{i}{\hbar}{\widehat h}(t_1)\Delta t}\varphi_{t_{0}}.
\end{equation}
Suppose the evolution of the state $\varphi$ of a particle is constrained to the classical space submanifold $M^{\sigma}_{3}$. The points $\varphi_{t_{0}}, \varphi_{t_{1}}, . . . $ belong then to the submanifold $M^{\sigma}_{3}$ and the steps can be identified with translations in the classical space. This is to say that for each $k$, the operator ${\widehat h}(t_{k})$ acts as the generator of translation by a vector ${\bf \xi}_{k}$ in $\R^3$, so that ${\widehat h}(t_{k})={\bf \xi}_{k}{\widehat {\bf p}}$, where ${\widehat {\bf p}}$ is the momentum operator. Because all operators of translation commute with each other, the equation (\ref{tN}) yields the following expression:
\begin{equation}
\label{tN1}
\varphi_{t_{N}}({\bf x})=\varphi_{t_{0}}(x-{\bf \xi}_{1}\Delta t-{\bf \xi}_{2}\Delta t- ... -{\bf \xi}_{N}\Delta t).
\end{equation}
That is, the initial state is simply translated by the vector 
\begin{equation}
\label{walk}
{\bf d}=\sum^{N}_{k=1}{\bf \xi}_{k}\Delta t
\end{equation}
in $\R^3$.
Now, the probability distribution of steps $-\frac{i}{\hbar} {\widehat h}(t_{k+1})\varphi_{t_{k}}$ in the tangent space $T_{\varphi_{k}}M^{\sigma}_{3}$ must be the conditional probability distribution of steps for the Hamiltonian satisfying ${\bf (RM)}$ under the condition that the steps take place in $T_{\varphi_{k}}M^{\sigma}_{3}$.
From the properties of the random matrix, it follows that ${\bf \xi}_k$ are independent and identically normally distributed random vectors, so that the equation (\ref{walk}) defines a random walk with Gaussian steps in $\R^3$. This is known to approximate the Brownian motion in $\R^3$ and yield the normal distribution of the position vector ${\bf d}$ at any time $t$, proving the claim.
\hfill$\square$

%Insert XX
Note that the theorem would not be true if the Hamiltonian were a random matrix in the Gaussian orthogonal ensemble in place of the unitary ensemble. In fact, the momentum operator in the proof is Hermitian but not orthogonal.
Now, the Gaussian unitary ensemble corresponds to systems that are not invariant under time reversal \cite{Dyson2, Wigner1}.
Therefore, the fact that we obtained in ${\bf (C)}$ a diffusion process on the submanifold $M^{\sigma}_{3}$ is tied to the fact that the Hamiltonian is not invariant under time reversal.

%end Insert X

By the extension of a random walk on $\R^3$ to the space of states $CP^{L_{2}}$, we understand a walk in $CP^{L_{2}}$ that satisfies  ${\bf (RM)}$ and that reduces to the original random walk on $\R^3$ when conditioned to stay on $M^{\sigma}_{3}$. 
From ${\bf (C)}$, we know that an extension of the walk with Gaussian steps on $\R^3$ exists. We claim that such an extension is {\it unique}. In fact, because the random walk conditioned to stay on $M^{\sigma}_{3}$ must be Gaussian, the entries of the matrix of the Hamiltonian associated with the directions tangent to $M^{\sigma}_{3}$ at a point are distributed normally. But the probability distribution of a single entry of the matrix defines the corresponding Gaussian unitary ensemble. Because of that, the random walk with Gaussian steps in $\R^3$ defines a unique random walk satisfying ${\bf (RM)}$. 
%
%Note also that a small change in the variation of the probability distribution function of steps of a Gaussian random walk on $\R^3 result in a small change of the probability distribution function of the corresponding random matrix with respect to obvious metric.
%
In what follows, the random walk of the state of a particle in ${\bf (RM)}$ will always be assumed to be this unique extension of the random walk with Gaussian steps that approximate the Brownian motion of the particle. We are free to make this choice as long as the resulting extension satisfies our needs. At the same time, this choice makes sense physically, because a displacement of the particle in $\R^3$ results in the like displacement of its state in $M^{\sigma}_{3}$, tying the two motions together. 

%in GUE defines the distribution of the matrix. Therefore, the statistical ensemble and the random walk that satisfies ${\bf (RM)}$ are uniquely determined.

The theorems ${\bf (A)}$ and ${\bf (C)}$ prove the consistency of the hypothesis ${\bf (EH)}$ for the deterministic Newtonian dynamics and the Brownian motion. 
Our goal now is to explain the origin of the constraint to the classical space and phase space submanifolds in the Hilbert space of states. For this, we need to establish properties of the random walk on the space of states in the conjecture ${\bf (RM)}$ and establish its applicability to the process of measurement.
 \begin{itquote}{\bf{(D)}}
The probability distribution of steps of the random walk specified in ${\bf (RM)}$ is isotropic and homogeneous. That is, for all initial states $\{\varphi\}$ in the space of states $CP^{H}$, 
the vector $d\varphi=-\frac{i}{\hbar}{\widehat h}\varphi dt$ is a normal random vector  in the tangent space $T_{\{\varphi\}}CP^{H}$, with spherical distribution.
The probability of reaching a certain state during the walk depends only on the distance between the initial and the final state.
\end{itquote}
{\it Proof.} To prove the theorem, note that because for any $t$ the matrix of ${\widehat h}$ is in GUE, the probability density function $P({\widehat h})$ of ${\widehat h}$ is invariant with respect to conjugations by unitary transformations. That is,  $P(U^{-1}{\widehat h}U)=P({\widehat h})$ for all unitary transformations $U$ acting in the Hilbert space of states.
%Indeed, for all unitary transformations $U$ acting in the Hilbert space of states, the probability density function $P({\widehat h})$ for ${\widehat h}$ is invariant with respect to conjugations by $U$, so that  $P(U^{-1}{\widehat h}U)=P({\widehat h})$. 
Also, for all unitary transformations $U$ that leave $\{\varphi\}$ unchanged and therefore all $U$ that act in the tangent space $T_{\{\varphi\}}CP^{H}$, we have
\begin{equation}
\label{v}
(U^{-1}{\widehat h}U \varphi, v)=({\widehat h}U \varphi, Uv)=({\widehat h}\varphi, Uv),
\end{equation}
where $v$ is a unit vector in $T_{\{\varphi\}}CP^{H}$. It follows that 
\begin{equation}
\label{vv}
P({\widehat h} \varphi, v)=P({\widehat h} \varphi, Uv),
\end{equation}
where $P$ is the probability density of the components of ${\widehat h}\varphi$ in the given directions.
By a proper choice of $U$, we can make $Uv$ to be an arbitrary unit vector in $T_{\{\varphi\}}CP^{H}$, proving the isotropy of the distribution.
On the other hand, for all unitary operators $V$ in $H$ and a unit vector $v$ in $T_{\{\varphi\}}CP^{H}$, we have
\begin{equation}
\label{vw}
P({\widehat h} \varphi, v)=P(V^{-1}{\widehat h}V \varphi, v)=P({\widehat h}V\varphi, Vv).
\end{equation}
Because $V\varphi$ is an arbitrary state and $Vv$ is in the tangent space $T_{\{V\varphi\}}CP^{H}$, we conclude with the help of (\ref{v}) that the probability density function is independent of the initial state of the system, proving the homogeneity of the distribution. The components of the vector ${\widehat h}\varphi dt$ are independent, because the entries of the matrix of ${\widehat h}$ are independent. It follows that $-\frac{i}{\hbar}{\widehat h}\varphi dt$ is a normal random vector with spherical distribution.
Finally, because different steps of the walk are independent, the probability of reaching a certain state during the walk may depend only on the distance between the initial and the final state.
\hfill$\square$

%Insert 11a

The fact that the components of the vector ${\widehat h}\varphi dt$ are independent signifies that the probability distribution of steps conditioned to take place in $M^{\sigma}_{3}$ is the same as the probability distribution of these steps without the condition. In particular, the random walk obtained in ${\bf (C)}$ is the usual random walk on $\R^3$, independent of the embedding of $\R^3$ into $CP^{L_{2}}$.
%End Insert 11a

Let us show now that 
%the version  ${\bf{(RM)}}$ of the BGS-conjecture 
conjecture ${\bf{(RM)}}$ provides a consistent approach to the process of observation  of the position of a single particle in the classical and quantum mechanics alike.
\begin{itquote}{\bf{(E)}}
The Schr{\"o}dinger evolution with the Hamiltonian that satisfies ${\bf (RM)}$
%represented by a family of random matrices in GUE parametrized by the time $t$ and independent for different values of $t$ 
models the measurement of the position of a particle in classical and quantum physics. Namely, under the constraint that the random walk that approximates the evolution of state takes place on the submanifold  $M^{\sigma}_{3}$, the probability distribution of the position random vector is normal. Without the constraint, 
the probability to find the initial state $\varphi_{0}$ at a point $\varphi$ is given by the Born rule. The transition from the initial to final state is time-irreversible.
\end{itquote}
{\it Proof.} In fact, according to  ${\bf (C)}$, the random walk that satisfies ${\bf (RM)}$ but is conditioned to stay on $M^{\sigma}_{3}$ is the random walk with Gaussian steps on $\R^3$. The latter random walk considered over the time interval of observation yields the normal distribution of the position random vector.  
%The fluctuations themselves can be attributed to random kicks exerted on the measured particle by the particles of the measuring device and the surroundings. 
On the other hand, according to ${\bf (D)}$, the probability $P$ of reaching a state $\varphi$ from the initial state $\varphi_0$ by means of a unconstrained random walk that satisfies ${\bf (RM)}$ depends only on the distance between the states. From ${\bf (B)}$, it then follows that $P$ is given by the Born rule: $P=|(\varphi, \varphi_0)|^2$.  Also, the choice of the Gaussian unitary ensemble corresponds to time-irreversible systems \cite{Wigner1}.
\null\nobreak \hfill$\square$

%Insert 12 from the theorem

Note that the normal distribution of the position agrees with observations in the macro world. It is also consistent with the central limit theorem applied to describe the cumulative effect of uncontrollable fluctuations from the mean in a series of measurement outcomes. It follows that the model satisfies the basic properties of observation of the position in classical and quantum mechanics. 

%End insert 12

Quantum mechanics does not explain how and when the deterministic Schr{\"o}dinger evolution is replaced by probabilistic evolution, whose outcomes obey the Born rule. It does not explain why macroscopic bodies are never observed in superpositions of position eigenstates.
The wave-function collapse models \cite{Bassi1,Bassi2} aim to answer these questions by introducing a non-linear stochastic modification of the Schr{\"o}dinger equation. It is assumed that the modified equation must be non-linear to be able to suppress superpositions of states. It is also assumed that the modified equation must vary with a change in the observed quantity. However, here the Born rule was derived from the linear Schr{\"o}dinger equation with the Hamiltonian satisfying  ${\bf (RM)}$. Moreover, it was derived for all observed quantities at once, without needed to change the equation of motion. Is there a contradiction?

The modified Schr{\"o}dinger equations in the collapse models make the state of the system converge to an eigenstate of the measured quantity, usually the position of a particle. An equation like that must break an arbitrary initial superposition of states, so it must be non-linear. Under the Schr{\"o}dinger evolution with the Hamiltonian satisfying the conjecture ${\bf (RM)}$, the state of the system does not converge to a position eigenstate. 
The equation does not suppress superpositions. It makes the state wander around the space of states and ensures that the probability of reaching a particular neighborhood in the space is given by the Born rule. That explains why the two approaches do not contradict each other.

We are now ready to investigate why the state of a macroscopic body is confined to the classical space submanifold of the space of states. 
The key to this is that the Brownian motion and the motion of state under a measurement are now derived from the same dynamics. We know that large particles do not experience a Brownian motion. That is because the total force acting on any such particle from the particles of the medium is nearly zero. As a result, in the absence of an external potential, the particle remains at rest in the medium. A similar mechanism explains why the state of a particle may be confined to the submanifold $M^{\sigma}_{3}$:
%Since the motion of state of the particle under a measurement is described by the same mechanism and the distribution of steps of the random walk is isotropic in the space of states, the state of the particle will be confined to $M^{\sigma}_3$ exactly under the same conditions that make the Brownian motion of the particle trivial.
\begin{itquote}{\bf{(F)}}
Suppose the Hamiltonian of a particle in the natural surroundings satisfies the conjecture ${\bf (RM)}$. Then, the state of the particle is constrained to the submanifold $M^{\sigma}_{3}$ precisely when the induced Brownian motion of the particle described in ${\bf (C)}$ vanishes. 
More precisely, under ${\bf (RM)}$, the boundary between the quantum and the classical occurs for particles that satisfy the following two conditions. The particles must be sufficiently large in size so that their Brownian motion in the natural surroundings is observable. At the same time, the particles  must not be too large, when the Brownian motion trivializes, the state initially in $M^{\sigma}_{3}$ becomes constrained to $M^{\sigma}_{3}$ and the Schr{\"o}dinger evolution becomes Newtonian motion.
\end{itquote}
{\it Proof.} Recall that the spaces $M^{\sigma}_{3}$ and $\R^3$ are metrically identical. Recall also that the random walk in ${\bf (RM)}$ is the unique extension of the random walk in $\R^3$ that approximates the Brownian motion of the particle in the surroundings.
Consider a particle whose initial state is in $M^{\sigma}_{3}$. 
%We know that when the unit of distance in $\R^3$ is equal to $2\sigma$, the classical space $\R^3$ is metrically identical to the manifold $M^{\sigma}_{3}$ with the induced metric. We also know that the displacement of a particle in $\R^3$ is physically represented by the corresponding displacement of state in $M^{\sigma}_{3}$. 
Suppose the particle is sufficiently large in size so that the Brownian motion of the particle in the natural surroundings is negligible. Therefore, the motion of state in the directions tangent to $M^{\sigma}_{3}$ is negligible. From ${\bf (D)}$, we know that for the Hamiltonian satisfying ${\bf (RM)}$, the probability distribution of steps of the random walk in the space of states of the particle is isotropic. Because the probability of steps in the directions tangent to $M^{\sigma}_{3}$ vanishes, the same holds true for any other direction tangent to the space of states at the same point. 
%It follows that all the terms in the decomposition  (\ref{decomposition}) for the state in the projective manifold $M^{\sigma}_{3,3}$ vanish. 
Therefore, in the absence of an external potential, the motion of state of the particle in the space of states is trivial. 
When an external potential is applied to the particle, the two middle terms in the decomposition (\ref{decomposition}) may appear. However, these terms can only contribute to the motion of state in the direction tangent to the classical phase space submanifold $M^{\sigma}_{3,3}$. Therefore, the state will remain constrained to the submanifold. The theorem ${\bf{(A)}}$ asserts then that the particle in such a state will move in accord with the Newtonian dynamics. 

On the contrary, suppose the particle is sufficiently small, but not too small, so that the interaction between the particle and the surroundings cannot be ignored and results in a noticeable Brownian motion of the particle. By the isotropy of the probability distribution of steps of the random walk of state, a displacement of the particle away from $M^{\sigma}_{3}$ is then equally likely. Such a displacement would mean that the particle is now in a superposition of states of different positions. If ${\bf a}$ is the initial position and ${\bf l}$ is the observed displacement of the particle in $M^{\sigma}_{3}$ during the measurement, then the states $g_{{\bf a}, \sigma}$ and $g_{{\bf a}+{\bf l}, \sigma}$ are distinguishable in the experiment. It means that the superpositions of these states can be observed as well, indicating that we are dealing with a quantum system.
\hfill$\square$

We see how the properties of the Hamiltonian in the conjecture ${\bf (RM)}$ are responsible for the fact that the state of a macroscopic particle driven by the Schr{\"o}dinger dynamics with this Hamiltonian is constrained to the classical phase space manifold. One might be tempted to say that the resulting ``freezing" of the state is in agreement with the quantum Zeno effect for the particle whose position is continuously measured.  However, the essential difference is that the result is derived here from the unitary evolution generated by a random Hamiltonian without ever needing to involve the projection operators.

As an example, 
%in the Appendix we
let us find the displacement of a particle of radius $1mm$ in still air in normal conditions. 
%Assuming the particle's position is measured by scattering and observing visible light, the displacement during the time interval of the measurement is found to be of the order of $10^{-12}m$. The corresponding displacement of state in the space of states is of the order of $10^{-7}rad$. These displacements are too small to be observed in the described experiment. The state of the particle remains ``frozen". The particle is constrained to $M^{\sigma}_{3}$ and no superpositions of states of a different position of the particle can be observed. 
%
%Insert 13
The estimate of the diffusion coefficient of a macroscopic spherical particle of radius $1mm$ in still air in normal conditions is based on the Stokes-Einstein relationship
\begin{equation}
D=\frac{k_{B}T}{6\pi \eta r},
\end{equation} 
where $D$ is the diffusion coefficient, $r$ is the radius of the particle, $\eta$ is the dynamic viscosity, $T$ is temperature of the medium and $k_{B}$ is the Boltzmann constant. Using the room temperature and the known value of dynamic viscosity $\eta \sim 10^{-5}N\cdot s/m^2$, we get $D \sim 10^{-12}m^2/s$. The variance of the $x$-coordinate of position of the particle is given by
$\overline{x^{2}}=2Dt$. If we scatter visible light off the particle to determine its position, the time interval of observation can be estimated to be as short as $10^{-13}s$. This amounts to the displacement of the order of $10^{-12}m$. The accuracy of measurement is limited by the wavelength $\lambda \sim 10^{-5}m$. The Fubini-Study distance between Gaussian states that are $10^{-12}m$ apart in $M^{\sigma}_{3}$ with $\sigma \sim 10^{-5}m$ is calculated via (\ref{mainO}) and is about $10^{-7}rad$. These displacements are too small to be observed in the described experiment. The state of the particle remains ``frozen". The particle is constrained to $M^{\sigma}_{3}$ and no superpositions of states of a different position of the particle can be observed.

Consider now a system consisting of a microscopic particle $P$ and a macroscopic device $D$ capable of measuring the position of $P$. The particle and the device form a two-particle system whose state belongs to the product Hilbert space $H=H_P \otimes H_D$. We have the following result.
%Change order 13a
\begin{itquote}{\bf{(G)}}
Suppose the initial state of the system consisting of a microscopic particle and a macroscopic device in the natural surroundings is a product state with the state of the device in the manifold $M^{\sigma}_{3}$. Suppose also that the conjecture ${\bf (RM)}$ holds true. Then, during the evolution, the state of the system remains in a product form with the state of macroscopic device confined to $M^{\sigma}_{3}$ and evolving classically. 
\end{itquote}
%End change order 13a
{\it Proof.} Consider the motion of the corresponding classical system. Let say, $D$ is a cloud chamber, small enough to be considered a material point, when treated classically. 
%Since the device is macroscopic, it interacts with the natural surroundings. It also interacts with the particle. 
%
The macroscopic device interacts with the surroundings. It also interacts with the particle. However, in classical physics, the effect of the particle on the device can be neglected. 
%Insert 14
It follows that we can apply ${\bf (RM)}$ to the device itself.
%End Insert 14
The interaction of the device with the surroundings results in a Brownian motion of the device. 
When the device is sufficiently large, its Brownian motion is trivial and the device is at rest in the lab system. By 
applying ${\bf (RM)}$, we conclude that the state of the device positioned initially in  $M^{\sigma}_{3}$ can be treated independently of the state of the particle and is at rest in the space of state $CP^{H_D}$. On the other hand, a small macroscopic particle placed in the medium of the chamber would undergo a Brownian motion. By 
applying ${\bf (RM)}$ to this case, we conclude that the state of the corresponding quantum system will perform a random walk in the space $CP^{H_P}$.  According to ${\bf (E)}$, the probability to find the particle at a particular point of $\R^3$ during the walk is given by the Born rule.

%By the previous discussion, the deterministic Schr{\"o}dinger evolution and the random walk that satisfies ${\bf (RM)}$ are unique extensions of their classical counterparts from a classical space submanifold to the space of states of the system. We conclude that the particle-device system initially in a product state must remain in a product state during the evolution.

%Insert 15
The conjecture ${\bf (RM)}$ is valid by assumption. Using the conjecture, we managed to predict the evolution of states of the device and the measured particle. This can only be the case when the state of the particle-device system remains separable throughout the evolution.
%End Insert 15
 Furthermore, according to ${\bf (F)}$, the state of the device will be confined to the submanifold $M^{\sigma}_{3}$ and in the presence of an external potential will be evolving in accord with the Newtonian dynamics. Under these conditions, the system remains in a product state with both factors being able to change. In particular, the position of the particle can be mechanically recorded with no entanglement between the particle and the device ever appearing or needing to appear in the process.
\hfill$\square$

\section{Implications for the notion of objective reality}

Theorem ${\bf (G)}$ provides an explanation of the issue of an apparent inconsistency of the description of quantum mechanics by macroscopic observers. The issue was pointed out first by Wigner \cite{Wigner2} and has picked up a significant interest in recent times.
%Insert 16 from the application paper
Wigner has demonstrated that the rules of quantum mechanics applied universally may lead to contradictory accounts of observations performed by two observers.  In his thought experiment \cite{Wigner2}, an observer (Wigner's friend) makes a measurement on a system and another observer (Wigner himself) observes the laboratory, where his friend performs the measurement. The contradiction is seen in the observers getting two different states after the friend has performed the measurement, but before the result is communicated to Wigner. This, as well as more recent thought experiments based on a similar scenario \cite{Renner}, raise the question of when the collapse of the state actually occur, what does the consciousness of the observers have to do with the collapse, and, ultimately, is there such a thing as an objective reality that all observers can agree on.
%End Insert 16

A possible reason for the contradiction in the accounts of Wigner and the friends in the Wigner's friend-type thought experiments is the assumption that a macroscopic observer or the whole lab may exist in a superposition of classical states. 
%Insert 17
Without this assumption the no-go result in \cite{Renner} doesn't hold and the original Wigner's friend paradox disappears. 
Note that the experiment reported in \cite{photonObserver}, in which the photons played the role of an observer is consistent with this conclusion. The reality as ``described" by microscopic particles is different than the one the macroscopic observers are aware of. 
Consequently, many researchers, including the authors of \cite{Renner} think that quantum mechanics is not directly applicable to macroscopic bodies. This is also the thinking behind the modifications of the Schr{\"o}dinger equation with the goal of suppressing the unobserved superpositions, and accounting for the transition to classicality and the measurement results  \cite{Bassi1,Bassi2}. 

The results proved in this paper offer an alternative solution to the problem. The Schr{\"o}dinger evolution may still be valid for macroscopic bodies, but we need to be careful in identifying the Hamiltonian of the corresponding system. Because of the unavoidable interaction of macroscopic bodies with the surroundings, the Hamiltonian cannot have the same form as before. 
%In fact, interaction of microscopic particles with the surroundings can usually be neglected in lab experiments.
In some ways, the macroscopic bodies are constantly ``measured" by the surroundings. At the same time, measuring a microscopic particle requires proper conditions. For instance, the needed conditions may be provided by a photographic plate, a cloud chamber, or a high-intensity radiation. It was assumed here that the Hamiltonian of a macroscopic particle in the natural surroundings and of a microscopic particle whose position is measured is represented by a random matrix that satisfies ${\bf (RM)}$. This assumption turned out to be sufficient to explain why no superpositions of classical states of macroscopic bodies are observed, to obtain irreversibility of a measurement and to derive the Born rule. 
Without superpositions of macroscopic observers and the lab, the observations by Wigner and the friends become consistent and objective reality is restored.

\end{document}